\documentclass{CSML}

\def\dOi{11(3:21)2015}
\lmcsheading%
{\dOi}
{1--14}
{}
{}
{Jan.~13, 2014}
{Sep.~25, 2015}
{}

\ACMCCS{[{\bf  Mathematics of computing}]: Mathematical
  analysis---Functional analysis; [{\bf Theory of computation}]:
  Semantics and reasoning---Program semantics---Denotational
  semantics}  

\usepackage{url,hyperref}

\newcommand{\R}{\mathbb{R}}

\newcommand{\oRp}{\overline{\mathbb{R}}_+}
\newcommand{\Rp}{{\mathbb{R}}_+}

\newcommand{\dbdownarrow}{\rlap{\raise.25ex\hbox{$\shortdownarrow$}}\raise-.25ex\hbox{$\shortdownarrow$}}
\newcommand{\dda}{\rlap{\raise.25ex\hbox{$\shortdownarrow$}}\raise-.25ex\hbox{$\shortdownarrow$}}
\newcommand{\dua}{\rlap{\raise-.25ex\hbox{$\shortuparrow$}}\raise.25ex\hbox{$\shortuparrow$}}
\newcommand{\dbuparrow}{\rlap{\raise-.25ex\hbox{$\shortuparrow$}}\raise.25ex\hbox{$\shortuparrow$}}
\newcommand{\diamondplus}{\mathop{\Diamond\mkern-13.9mu\raise.22ex\hbox{$+$}}}
\newcommand{\diamonddot}{\mathop{\Diamond\mkern-9.5mu\raise.2ex\hbox{$\cdot$}}\,}

\newcommand{\dsup}{\mathop{\bigvee{}^{^{\,\makebox[0pt]{$\scriptstyle\uparrow$}}}}}

\newcommand{\cH}{\mathcal{H}}
\newcommand{\cO}{\mathcal{O}}

\newcommand{\cL}{\mathcal{L}}

\newcommand{\cV}{\mathcal{V}}

\newcommand{\da}{\mathord{\downarrow}}
\newcommand{\ua}{\mathord{\uparrow}}

\newcommand{\display}{:}
 \theoremstyle{definition}\newtheorem{question}[thm]{Question}

\newcommand\cone{\mathsf{cone}}

\begin{document}

\title[Weak upper topologies]{Weak upper topologies and duality for cones}
 \author[K.~Keimel]{Klaus Keimel}
 \address{Fachbereich Mathematik\\ Technische Universit\"at
  Darmstadt\\ Schlo\ss gartenstra\ss e~7, D--64289~Darmstadt, Germany}
 \email{keimel@mathematik.tu-darmstadt.de}
 \thanks{Supported by DFG (Deutsche Forschungsgemeinschaft)}

\maketitle

\begin{abstract}
In functional analysis it is well known that every linear functional
defined on the dual of a locally convex vector space which is continuous for
the weak$^*$ topology is the evaluation at a uniquely determined
point of the given vector space. 
M.~Schr\"oder and A.~Simpson have obtained a similar result for
lower semicontinuous linear functionals on the cone of all
Scott-continuous valuations on a topological space endowed with the
weak$^*$upper topology, an asymmetric version of the weak$^*$
topology. This result has given rise to several proofs, originally by 
Schr\"oder and Simpson themselves and, more recently, by the author of these
Notes and by J. Goubault-Larrecq. The proofs developed from very
technical arguments to more and more conceptual ones. The present Note
continues on this line, presenting a conceptual approach inspired by
classical functional analysis which may prove useful in other situations. 
\end{abstract}

\section{Introduction}

Recently, a theorem announced by Schr\"oder and Simpson in \cite{SS} and
presented with a proof in \cite{SS1} has attracted some
attention.  Before stating this result we need some preparations.

Let $X$ be an arbitrary topological space. The emphasis is on non-Hausdorff 
spaces like directed complete partially ordered sets (= dcpos) with
the Scott topology (see \cite{gierz03}) and quotients 
of countably based spaces (qcb-spaces, see \cite{BSS}). But, of course,
compact and locally compact  
Hausdorff spaces are classical examples. We consider functions
defined on $X$ with values in 
$\oRp$, the set $\Rp$ of nonnegative real numbers extended by
$+\infty$ as top element. Such  functions $f$ are \emph{lower
  semicontinuous} if, for every $r\in\Rp$, the set $\{x\in X\mid
f(x)>r\}$ is open in $X$. 
We denote by
$\cL X$ the set of all lower semicontinuous functions $f\colon X\to
\oRp$ with the pointwise defined order. The pointwise supremum of any 
family of lower semicontinuous functions is lower semicontinuous so
that $\cL X$ is a complete lattice. The pointwise defined sum $f+g$ of two
lower semicontinuous functions $f,g$ as well as the scalar multiple
$rf$ for $r\in \Rp$ are also lower semicontinuous. 

The maps $\mu\colon \cL X\to \oRp$ which are linear and
Scott-continuous, that is, those 
satisfying \[\mu(f+g) = \mu(f)+\mu(g),\ \ \mu(rf)=r\mu(f),\ \
\mu(\dsup_i f_i) = \dsup_i\mu(f_i)\] for all $f,g\in\cL X$, $r\in\Rp$ and  
for every directed family
$(f_i)_i$ in $\cL X$, form the \emph{valuation powerdomain}
$\mathcal V X$. On $\cV X$ one also has a pointwise defined order,
addition and multiplication by nonnegative real numbers. We endow $\cV
X$ with the weak$^*$upper topology, that is, the coarsest topology
such that the evaluation maps $\mu\mapsto \mu(f)\colon \cV X\to\oRp$
remain lower semicontinuous for all $f\in \cL X$. The Schr\"oder-Simpson
theorem says:

\begin{thm}\label{SS} 
For every linear functional $\varphi\colon \cV X\to\oRp$ which is
lower semicontinuous with respect to the weak$^*$upper topology, there
is an $f\in\cL X$ such that $\varphi(\mu)=\mu(f)$ for all $\mu\in\cV X$.
\end{thm}  

This theorem sounds exactly like a simple general fact from classical functional
analysis (see, e.g., \cite[Chapter IV, 1.2]{Sch}): \emph{Let $V$ be a
  topological 
vector space over the field of real numbers and $V^*$ the dual vector
space of all continuous linear functionals $\mu\colon V\to \R$, then
every linear functional $\varphi$ defined on the 
topological dual $V^*$  which is continuous with respect to the weak$^*$
topology is the evaluation at some $f\in V$, that is, $\varphi(\mu)=
\mu(f)$ for every $\mu\in V^*$.} Here the weak$^*$ topology is the
coarsest topology on $V^*$ such that all the evaluation maps
$\mu\to\mu(f)\colon V^*\to\R$, $f\in V$, remain continuous. (Here $\R$
is meant to carry the usual Hausdorff topology). The idea behind the quite
simple proof of this classical
result is to reduce the question to the finite dimensional case, where
it is trivial. 

The Schr\"oder-Simpson Theorem is a
kind of asymmetric version of this classical fact in a special
situation. Its original proof was technical and involved. 
A new proof in a conceptual framework was published in
\cite{keimel12}. A short proof by J.~Goubault-Larrecq \cite{GL} has appeared
just recently. Let us remark that the Schr\"oder-Simpson
theorem has a predecessor in Domain Theory: If $X$ is a continuous
dcpo (with the Scott topology), then $\cV X$ is a continuous dcpo,
too, and every 
Scott-continuous linear functional on $\cV X$ is the evaluation at
some $f\in\cL X$ (see C. Jones \cite{J} and Kirch \cite{Ki}). For a continuous
dcpo $X$, the Scott topology on $\cV X$ agrees with the weak$^*$upper
topology as Kirch \cite{Ki} has shown, a fact that is now longer true
for more general spaces. Thus, the Schr\"oder-Simpson Theorem is a far
reaching generalization. 

The proofs of the Schr\"oder-Simpson theorem due to Keimel and
Goubault-Larrecq are based on two lemmas that hold for every
lower semicontinuous linear 
functional $\varphi$ from the valuation powerdomain $\mathcal
V X$ with the weak$^*$upper topology into $\oRp$, where $X$ is any
topological space\display

\begin{lem}{\rm  (Lemma 2 in \cite{GL}, Theorem 5.3 in
    \cite{keimel12})} \label{lem:2} 
There is a family $(f_i)_i$ of functions in $\cL X$ such that
$\varphi(\mu)=\sup_i\mu(f_i)$ for every $\mu\in\cV X$. 
\end{lem}

\begin{lem} {\rm (Lemma 1 in \cite{GL}, Lemma 5.5 in
    \cite{keimel12})} \label{lem:1} 
The set of all $f\in\cL X$ such that $\varphi(\mu)\geq \mu(f)$ for all
$\mu\in\cV X$ is directed. 
\end{lem}

The Schr\"oder-Simpson Theorem \ref{SS} can now be deduced as follows:
Given a lower semicontinuous linear $\varphi\colon \cV X\to \oRp$,
Lemma \ref{lem:2} gives us a family of $(f_i)_i$ of functions in $\cL
X$ such that $\varphi(\mu)=\sup_i\mu(f_i)$ for all $\mu\in\cV X$.  
By Lemma \ref{lem:1} the family $(f_i)_i$ can be chosen to be directed
and we let $f=\sup_i f_i$. Since the valuations $\mu\in \cV X$ are
Scott-continuous on $\cL X$, they preserve directed suprema and we
obtain that $\varphi(\mu)=\sup_i\mu(f_i) =\mu(\sup_i f_i)=\mu(f)$ for
all $\mu\in \cV X$.  In fact, it suffices to find a subcone $C$ of
$\cL X$ and to prove the two lemmas above for subfamilies of $C$, for
example for the subcone of step functions.   

It is desirable to prove an analogue of the Schr\"oder-Simpson Theorem
for situations different from the valuation powerdomain.  
For the moment I do not know how to generalize Lemma \ref{lem:1}. For
Lemma \ref{lem:2}, a general conceptual argument had been developed in
\cite[Section 5.1]{keimel12} by reducing the problem to a situation close to the
classical vector space situation; but this argument 
needed some quite heavy background from the theory of quasi-uniform
locally convex cones.  Goubault-Larrecq found a simple direct argument
for the proof of Lemma \ref{lem:2} for the special 
situation of the valuation powerdomain.  
In this note we present a direct approach to Lemma \ref{lem:2} based on
Goubault-Larrecq's idea but in an appropriate generality.
It uses the classical idea to reduce the
problem to the finite dimensional case.  
   
Let us recall the motivation behind all of this.
The Schr\"oder-Simpson theorem concerns the probabilistic
powerdomain over arbitrary topological spaces used in semantics for
modelling probabilistic phenomena in programming. Goubault-Larrecq  \cite{GL12}
recently has applied the result in his work on modelling probabilistic
and ordinary nondeterminism simultaneously.

\section{Cones and separation}

The Schr\"oder-Simpson Theorem cannot be proved with methods from
classical functional analysis, since the real-valued lower
semicontinuous functions on a topological space $X$ do not form a
vector space. For a lower semicontinuous function $f\colon X\to \R$,
its negative $-f$ is not necessarily lower semicontinuous.  Similarly, we
are in an asymmetric situation for the valuations, those being
Scott-continuous linear functionals. We therefore work with cones, a
weakening of the notion of a vector space by restricting
scalar multiplication to nonnegative reals and not requiring the
existence of additive inverses. This offers the advantage of including
infinity. 
In this section we provide some basic concepts. The connection to our
goal will appear only later.

We denote by $\Rp$ the set of nonnegative real numbers with the usual
 linear order, addition and multiplication. The letters $r,s,t, \dots$
 will always denote nonnegative reals. Further
 $$ \oRp=\Rp\cup\{+\infty\}$$ 
denotes the nonnegative reals extended by $+\infty$ as a top element.
 We also extend
addition and multiplication from $\Rp$ to $\oRp$ by defining $r+\infty
=+\infty+r =+\infty$ for all $r\in\oRp$ and
\[r\cdot(+\infty)=+\infty\cdot r =
        \begin{cases} +\infty & \mbox{ for } r>0\\
                      0       & \mbox{ for } r=0
        \end{cases}  \]

\begin{defi}\label{cone}{\rm
A \emph{cone} is a commutative monoid $C$ carrying a
scalar multiplication by nonnegative real numbers satisfying the same axioms
as for vector spaces; that is, $C$ is endowed with an addition
$(x,y)\mapsto x+y\colon C\times C\to C$ and a neutral element $0$
satisfying\display 
\[\begin{array}{rcl}
x+(y+z)&=&(x+y)+z\\
x+y&=&y+x\\
x+0&=&x
\end{array}\]
  and with a scalar multiplication
$(r,x)\mapsto r\cdot x\colon \Rp\times C\to C$ satisfying\display
\[\begin{array}{rcl}
r\cdot (x+y)&=&r\cdot x+ r\cdot y\\
(r+s)\cdot x&=&r\cdot x+ s\cdot x\\
(rs)\cdot x&=&r\cdot (s\cdot x)\\
 1\cdot x&=&x\\
0\cdot x&=&0
\end{array}
\]
for all $x,y,z\in C$ and all $r,s\in\Rp$.
An \emph{ordered cone} is a cone $C$ endowed with a partial order
$\leq$, such that addition and multiplication by fixed scalars $r\in
\R_+$ are order preserving, that is, for all $x,y,z \in C$ and all
$r\in \R_+$\display 
\[x\leq y \Longrightarrow x+z\leq y+z \mbox{ and } r\cdot x \leq
r\cdot y\]}
\end{defi}

Unlike for vector spaces, addition in cones need 
not satisfy the cancellation property, and cones need not
be embeddable in vector spaces. For example $\oRp =
\Rp\cup\{+\infty\}$ is an ordered cone 
that is not embeddable in a vector space. The same holds for finite powers
$\oRp^{\, n}$ and for infinite powers $\oRp^{\, I}$ with the pointwise
defined order $x\leq y$ iff $x_i\leq y_i$ for all indices $i$.  
Maps $\varphi$ from a cone $C$ into  $\oRp$ are simply
referred to as \emph{functionals}. Thus, functionals
are allowed to have the value $+\infty$.  
On any set of functionals we consider the pointwise order
$\varphi\leq \psi $ if $\varphi(x)\leq\psi(x)$ for all $x\in C$.

\begin{defi}{\rm
Let $C$ be a cone. A functional\footnote{We simply call homogeneous what is usually
  called positively homogeneous. Previously authors always supposed
  $\varphi(0)=0$ for homogeneous functionals; note that we omit this
  requirement.} $\varphi\colon C\to \oRp$ is
called
\\[2mm]
\begin{tabular}{lrcll}
\emph{homogeneous}, if\display &
$\varphi(r\cdot a)$ &$=$& $r\cdot \varphi(a)$& for all  $a\in C$ and
all $r\in\Rp$\, ,\\[2mm]
\emph{additive}, if\display&
$\varphi(a+b)$&=&$\varphi(a) + \varphi(b)$& for all  $a,b\in C$\, ,\\[2mm]
{\em superadditive} if\display&
$\varphi(a+b)$&$\geq$&$\varphi(a) + \varphi(b)$& for all  $a,b\in C\, ,$\\[2mm]
{\em subadditive} if\display& 
$\varphi(a+b)$& $\leq$ &$\varphi(a)+\varphi(b)$& for all  $a,b\in C \, .$
\end{tabular}\\[2mm]
We say that $\varphi$ is \emph{sublinear} (\emph{superlinear},
\emph{linear}, respectively), if it is  
homogeneous and subadditive (superadditive,
additive, respectively).}
\end{defi}

As in real vector spaces, we have a
notion of convexity\display

\begin{defi}\label{def:convex}{\rm
A subset $K$ of a cone $C$ is \emph{convex} if, for all $a,b\in K$, the
convex combinations $ra + (1-r)b$ belong to $K$ for all real
number $r$ with $0\leq r\leq 1$.  }
\end{defi}

Convex sets in cones may look different than those in vector
spaces because of the presence of infinite elements. Indeed, in
$\oRp$ not only intervals $\mathbb I$ are convex but also the sets 
$\mathbb I \cup\{+\infty\}$.


To every functional $\varphi\colon C\to\oRp$ we assign the sets 
\[U_\varphi = \{x\in C\mid \varphi(x)> 1\},\ \ A_\varphi =\{x\in C\mid
\varphi(x)\leq 1\} \]  which are complements one of the other. These
sets reflect properties of the functionals\display

\begin{rem}\label{rem:mink}
For homogeneous functionals $\varphi$ and $\psi$ on a cone $C$ we have\display

(a) $\varphi\leq \psi$ if and only if $U_\varphi\subseteq U_\psi$ if
and only if $A_\varphi\supseteq A_\psi$. 

\noindent
{\rm Indeed, $\psi\leq\varphi$ iff, for all $0<r<+\infty$, $\psi(x)\leq r
\implies \varphi(x)\leq r$.  For $0<r<+\infty$,  $\psi(x)\leq r
\implies \varphi(x)\leq r$ iff
$\psi(\frac{x}{r})\leq 1 \implies \varphi(\frac{x}{r})\leq 1$ iff  
$\frac{x}{r}\in A_\psi \implies \frac{x}{r}\in A_\varphi$ iff $x\in
rA_\psi \implies x\in rA_\varphi$ iff $rA_\psi \subseteq  rA_\varphi$
iff $A_\psi \subseteq  A_\varphi$.}

(b) If $\varphi$ is sublinear then $A_\varphi$ is convex.

\noindent
{\rm Indeed, if $\varphi$ is sublinear and $a,b$ are elements in
  $A_\varphi$, then $\varphi(ra+(1-r)b)\leq \varphi(ra)
+\varphi((1-r)b) = r\varphi(a) +(1-r)\varphi(b) \leq 1$, whence
$ra+(1-r)b\in A_\varphi$.}

(c) If $\psi$ is superlinear then $U_\psi$ is convex. 
\end{rem}

The following technical result is the key to our main result\display

\begin{lem}\label{lem:sep}
Let $V=\{(y_1,\dots,y_n)\in\oRp^n\mid y_i> 1 \mbox{ for all }
i=1,\dots,n\}$ and $K$ be a convex subset in $\oRp^n$ disjoint from
$V
$. Then there are nonnegative real numbers 
$a_1,\dots,a_n$ such that  $\sum_{i=1}^n a_i =1$ and 
\[\sum_i a_i x_i\leq 1 < \sum_i a_i y_i\ \ \  \ \mbox{ for all
  }(x_1,\dots,x_n)\in K, (y_1,\dots,y_n)\in V\]
\end{lem}

\proof Let $V$ and $K$ be as in the statement of the lemma.
The convex hull of $K\cup [0,1]^n$ does not meet $V$. Indeed, if $x\in K$, then $x_i\leq 1$ for some coordinate $x_i$ of $x$; as $y_i\leq 1$ for every coordinate of $y\in[0,1]^n$, the $i$-th coordinate of every convex combination of $x$ and $y$ is $\leq 1$.  As $V$ is an upper set, the lower set $L$ of all elements $z\in \oRp^n$ that are below some element in the convex hull of $K\cup[0,1]^n$ does not meet $V$, and $L$ is convex, too. 

We now consider the  convex subset $L'=L\cap \R^n$ and the convex open
subset $V'=V\cap\R^n$ of $\R^n$ which are disjoint. By a standard
separation theorem (see the remark below), there is a linear
functional $f\colon \R^n\to \R$ such that $f(x)\leq 1< f(y)$ for all
$x\in L'$ and all $y\in V'$. As for every linear functional on $\R^n$,
there are real numbers $a_1,\dots,a_n$ such that
$f(x_1,\dots,x_n)=\sum_i a_ix_i$ for
every $x=(x_1,\dots,x_n)\in\R^n$. Let us show that $a_i\geq 0$ for
every $i$ and $\sum_i a_i=1$.  

Since $(1,1,\dots,1)\in L'$, we have  $\sum_i a_i =f(1,1,\dots,1)\leq
1$. Since $(1+\varepsilon,\dots,1+\varepsilon)\in V'$ for every
$\varepsilon >0$, we have $\sum_i a_i +n\varepsilon =
f(1+\varepsilon,\dots,1+\varepsilon)> 1$ for all $\varepsilon > 0$,
whence $\sum_i a_i \geq 1$. We finally show that $a_1\geq 0$. Since
the vector $(2+\varepsilon,1+\varepsilon,\dots, 
1+\varepsilon)$ belongs to $V'$ for every $\varepsilon >0$, we have $1
< f(2+\varepsilon,1+\varepsilon,\dots, 1+\varepsilon)=a_1 + \sum_{i}^n
a_i +n\varepsilon=a_1+1+n\varepsilon$, that is, $0<a_1+n\varepsilon$
for every $\varepsilon >0$, which implies $a_1\geq 0$. Similarly,
$a_i\geq 0$ for all $i$.    

The restriction of $f$ to the positive cone $\Rp^n$ has a unique
Scott-continuous extension to $\oRp^n$ also given by $\sum_i a_ix_i$
for all $x=(x_1,\dots,x_n)\in \oRp^n$ which has the desired separation
property for $K$ and $V$. 
\qed

\begin{rem}
In the proof of the preceding Lemma \ref{lem:sep} we have used a
standard Separation Theorem which can be proved for arbitrary
topological vector spaces $E$ over the reals (see, e.g., \cite[Theorem
3.4(a)]{Ru}): \emph{For every open convex 
subset $V'$ and every convex subset $L'$ disjoint from $V'$ there are a
continuous linear functional $f\colon E\to \R$ and a real number $r$
such that $f(x)\leq r< f(y)$ for every $x\in L'$ and every $y\in
V'$. Since $r\neq 0$ in the situation in the proof above, one may
choose $r=1$ by replacing $f$ by $\frac{1}{r}f$. }

The proof of this standard Separation Theorem 
uses the axiom of choice. We only need this Separation Theorem for
finite dimensional real vector spaces. For this special case, there
exist constructive proofs. One of the referees for this paper has
worked out such a constructive proof.  One may also use
a separation theorem for convex subsets in finite dimensional vector
spaces (see Aliprantis and Border [1, Theorem 7.35, p. 279]) often
attributed to Minkowski. In \cite{B}, K. C. Border 
discusses the same result under the name of 'Finite Dimensional Separating
Hyperplane Theorem 11', and Border's proof looks constructive,
as far as I can see. \emph{Thus, our Lemma \ref{lem:sep} has a constructive
proof, not using the axiom of choice or any weak form thereof, and the
same holds for the following proposition.} 

I am indebted to one of the referees for a detailed discussion on the
issue of a constructive proof. Moreover, the proof of Lemma
\ref{lem:sep} has been simplified by a suggestion of this referee, namely to
replace $K$ by the convex hull of $K$ with the cube $[0,1]^n$.\qed
\end{rem}

The following key result uses the reduction to the finite dimensional
case as indicated in the Introduction\display  

\begin{prop}\label{key}
Consider a cone $D$ together with a sublinear functional
$\varphi\colon D\to\oRp$ and finitely many linear functionals
$\psi_i\colon D\to\oRp$, $i=1,\dots,n,$ such 
that $\psi_1\wedge\dots\wedge\psi_n\leq \varphi$. Then there are
nonnegative real numbers $a_1,\dots,a_n$ such that $\sum_i a_i=1$ and 
$\psi_1\wedge\dots\wedge\psi_n\leq 
\sum_{i=1}^na_i\psi_i \leq \varphi$. 
\end{prop}

\proof
Let $A=A_\varphi$ be the set of all $x\in D$ such that $\varphi(x)\leq 1$. Since
$\varphi$ is supposed to be sublinear, $A$ is convex. Further let  
$U$ be the set of all $x\in D$ such that
$(\psi_1\wedge\dots\wedge\psi_n)(x)=\min_i\psi_i(x)>1$. 
Consider the map $\Psi\colon D\to\oRp^n$ defined by 
$\Psi(x)=(\psi_1(x),\dots,\psi_n(x))$. This map $\Psi$ is
linear, since all the $\psi_i$ are linear 
. For $x\in U$ we 
have $\psi_i(x)>1$ for every $i$, that is
$\Psi(x)\in V=\{(x_1,\dots,x_n)\in\oRp\mid x_i>1, i=1,\dots,n
\}$. Since $\Psi$ is linear, the image $K=\Psi(A)$ of $A$ is
convex. For $a\in A$, 
we have $(\psi_1\wedge\dots\wedge\psi_n)(a)\leq\varphi(a)\leq
1$, whence $(\psi_1(a),\dots,\psi_n(a))\not\in V$. Thus, 
$K=\Psi(A)$ is a convex subset of $\oRp^n$ disjoint from 
$V$. 
By the Separation Lemma \ref{lem:sep} there are
nonnegative real numbers $a_1,\dots,a_n$ 
such that $\sum_i a_i = 1$ and $\sum_ia_ic_i\leq 1<\sum_ia_id_i$
for all $c\in K$ and all $d\in V$. We denote by $G\colon \oRp^n\to\oRp$
the linear map $G(x)=\sum_i a_ix_i$. Then,
on one hand, $G\circ \Psi$ is a 
linear functional on $D$. If $(\psi_1\wedge\dots\wedge \psi_n)(x)>1$,
then  $x\in U$, hence $\Psi(x)\in V$ which implies
$G(\Psi(x))>1$, whence $\psi_1\wedge\dots\wedge \psi_n \leq
G\circ\Psi$ by Remark \ref{rem:mink}(a).  
On the other hand,
if $\varphi(a)\leq 1$, then $a\in A$, whence  $G(\Psi(a))\leq 
1$; thus $G\circ\Psi\leq \varphi$ again by Remark \ref{rem:mink}(a).
Noticing that, $G(\Psi(x))=\sum_{i=1}^n
a_i\psi_i(x)$, that is, $G\circ \Psi =\sum_{i=1}^n a_i\psi_i$, we have
the desired result.        
\qed

In the above lemma, the maps $\psi_i$ are supposed to be linear; so the linear
combination $\sum_ia_i\psi_i$ is linear, too. In particular, we
have interpolated a linear functional between $\varphi$ and
$\psi_1\wedge\dots\wedge\psi_n$. Under the hypotheses of the
preceding lemma, there are general separation theorems that yield directly
 the existence   
of a linear functional $\psi$ between the sublinear functional $\varphi$
and the superlinear functional 
$\psi_1\wedge\dots\wedge\psi_n$. Our Lemma above gives us more
information as it tells us that $\psi$ may be chosen to be a convex
combination of the $\psi_i$ and this property will be crucial in
the following. 

One may prove the preceding lemma also under the weaker hypothesis that
the maps $\psi_i$ are only superlinear and not linear. But we have no
use for this generalization.

\section{Semitopological cones and lower semicontinuous functionals}

Every partially ordered set $P$ can be endowed with a simple
topology, the \emph{upper topology} $\nu$: a subbasis for the
closed sets is given by the principal ideals $\da x=\{y\in P\mid y\leq
x\}$, $x\in P$. 

If not specified otherwise, we
will use the upper topology on $\Rp$ and on  $\oRp$. The proper
nonempty open subsets are simply the open upper intervals
$]r,+\infty[=\{x\mid x>r\}$ on $\Rp$ and  
the open upper intervals $]r,+\infty]$ on $\oRp$, where $r\in\Rp$.

This asymmetric upper topology on $\oRp$ is appropriate to talk about
lower semicontinuity. 
In agreement with classical analysis, for an arbitrary topological
space $X$, we 
call {\em lower semicontinuous}\footnote{It is
somewhat unfortunate that those functions are called \emph{lower}
semicontinuous which are continuous with respect to the \emph{upper}
topology. But we do not want to deviate from the terminology in
classical analysis and the one adopted in \cite{gierz03}.} 
those functions $f\colon X\to \oRp$
which are continuous with respect to the upper topology on $\oRp$.
Extended addition and multiplication are lower semicontinuous as maps
$\oRp\times\oRp \to\oRp$ also at infinity. 

For topological spaces $X,Y,Z$, a function $f\colon X\times Y\to Z$
will be said to be \emph{separately continuous} if, for each fixed
$x\in X$, the function $y\mapsto f(x,y)\colon Y\to Z$ and,
for each fixed $y\in Y$, the function $x\mapsto f(x,y)\colon X\to Z$
is continuous. We say that $f$ is \emph{jointly 
  continuous} if it is continuous for the product topology on
$X\times Y$. 

\begin{defi}{\rm
A \emph{semitopological cone} is cone equipped with a topology
satisfying the T$_0$ separation axiom such that
addition  $(x,y)\mapsto x+y\colon C\times C\to C$ and scalar
multiplication $(r,x)\mapsto rx\colon \Rp\times C\to C$ are separately
continuous. If these maps are jointly continuous, we have a 
\emph{topological cone}.}  
\end{defi}

\begin{exa}\label{ex:rn}
For a nonnegative integer $n$, clearly $\oRp^n$ is a cone ordered by the
coordinatewise order. 
The upper topology on $\oRp^n$ is the product of the upper
topology on $\oRp$. Addition and extended scalar multiplication are
continuous for the upper topology on $\oRp^n$ so that $\oRp^n$ is a
topological cone.
The same holds for infinite powers $\oRp^I$.
\end{exa}

Any topological T$_0$ space $X$ comes with an intrinsic partial order, the
\emph{specialization preorder} defined
by $x\leq y$ if the element $x$ is contained in the closure of the
singleton $\{y\}$ or, 
equivalently, if every open set containing $x$ also contains
$y$. In this paper, an order on a topological space will always be the
specialization order. 
Open sets are upper (= saturated) sets and closed sets are lower sets with
respect to the specialization preorder. The saturation $\ua A$ of a
subset $A$ can also be 
characterized to be the intersection of the open sets containing $A$.   
In Hausdorff spaces, the specialization order is trivial.
On $\oRp$, the specialization order for the upper topology agrees with the 
usual linear order.  Continuous maps between
topological T$_0$ spaces preserve the respective specialization
preorders. For more details, see e.g. \cite{gierz03}, Section 0-5.

As we use the upper topology on $\Rp$, a semitopological cone
$C$ cannot satisfy the Hausdorff separation property:
Since continuous maps preserve the respective specialization
preorders, the continuity of scalar multiplication
implies firstly that, for every $a\in C$, the map
$r\mapsto ra\colon\R_+ \to C$ is order preserving, that is, the
rays in the cone are linearly 
ordered, and secondly that the
cone is pointed; indeed, $0=0\cdot a\leq 1\cdot a=a$ for all elements
$a$ of the cone. In particular, the 
specialization order on $C$ is nontrivial and the topology is far from
being Hausdorff except for the trivial case $C=\{0\}$. 
As continuous maps preserve the specialization order, we
conclude that \emph{every semitopological cone is an ordered  cone
  with bottom element $0$}.

Let $D$ be a semitopological cone. There is a close relation between
classes of open subsets of $D$ and classes
of lower semicontinuous functionals on $D$. 

An arbitrary functional $\varphi\colon D\to \oRp$ is lower
semicontinuous if and only if, for every $r\in \Rp$, the set $\{x\in
D\mid \varphi(x)>r\}$ is open. For a homogeneous functional this
simplifies: A homogeneous functional
 is lower semicontinuous\footnote{Homogeneous
  lower semicontinuous functionals on cones of lower semicontinuous maps are 
  called \emph{previsions} by Goubault-Larrecq.}
 if and only if the set  
$$U_\varphi = \{y\in D\mid \varphi(y)>1\}$$
is open. Since $\varphi(0)=0$, the open set $U_\varphi$ does not
contain $0$. Conversely, for any open subset $U\subseteq D$ containing
$0$, the \emph{upper Minkowski functional}
$$\varphi_U(y)=\sup\{r>0\mid  y\in r\cdot U\}$$ is lower semicontinuous and
homogeneous;  it is understood that $\varphi_U(x) = 0$, if
$x\not\in rU$ for all $r>0$.

For any nonempty finite family $\varphi_1,\dots,\varphi_n$ of homogeneous lower
semicontinuous functions 
the pointwise infimum \[(\varphi_1\wedge\dots\wedge \varphi_n)(x) =
\min(\varphi_1(x),\dots,\varphi_n(x))\] is again
homogeneous and lower semicontinuous. The same holds for the pointwise
 supremum $\big(\bigvee_i\varphi_i\big)(x)=\sup_i\varphi_i(x)$ of an
 arbitrary family $(\varphi_i)_i$ of 
homogeneous lower semicontinuous functionals.

%

We have the following properties sharpening Remark \ref{rem:mink} (see
\cite[Propositions 7.4 and 7.5]{keimel08} and \cite{Plo})\display

\begin{lem}\label{lem:hom}\hfill
\begin{enumerate}[label=\({\alph*}]
\item The assignment $\varphi\mapsto U_\varphi$ establishes an order
isomorphism between the collection $\cH X$ of all lower semicontinuous
homogeneous functionals $\varphi \colon D\to\oRp$ ordered pointwise
and the collection $\cO^* X$ of all proper open subsets $U$ of $D$
ordered by inclusion; the inverse map is given by assigning to every
$U\in\cO^* X$ its Minkowski functional  $\varphi_U(x)
$.  
For homogeneous  lower semicontinuous functionals we have\display
\begin{eqnarray}
\varphi\leq\psi &\iff& U_\varphi \subseteq U_\psi,\label{1}\\ 
U_{\varphi_1}\cap\dots \cap U_{\varphi_n}
&=&U_{\varphi_1\wedge\dots\wedge \varphi_n},\label{2}\\ \bigcup_i U_{\varphi_i}
&=& U_{(\bigvee_i \varphi_i)}. \label{3}
\end{eqnarray}    

\item A homogeneous  functional $\varphi$ is superlinear if and only if
the corresponding open set $U_\varphi$ is convex. 

\item A homogeneous functional $\varphi$ is sublinear if and only if the
complement $A_\varphi= D\setminus U_\varphi$ of the corresponding open set is
convex.\qed
\end{enumerate}
\end{lem}

\section{Weak upper topologies and the main result} 

We fix a \emph{dual pair of cones} which, by definition,
consists of two cones $C$ and $D$ together with
a bilinear map $(x,y)\mapsto\langle x,y\rangle\colon C\times
D\to\oRp$. Bilinearity means that for each $x\in C$, the map
$$\widehat x = (y\mapsto \langle x,y\rangle)\colon D\to \oRp$$
is linear and similarly for fixed $y\in D$. We will always suppose this
bilinear map to be non-singular, that 
is, for any two different elements $y\neq y'\in D$ there is an $x\in C$
such that $\langle x,y\rangle\neq \langle x,y'\rangle$.

As before, we endow $\oRp$ with the \emph{upper} topology. 
We consider the coarsest topology $w(D,C)$ on $D$
rendering lower semicontinuous the maps
$\widehat x = (y\mapsto\langle x,y\rangle) \colon 
D\to \oRp$ for all $x\in C$. We call it the \emph{weak upper topology}
on $D$. A subbasis of this topology is given by the sets 
$$U_{x} =\{y\in D\mid \widehat x(y)=\langle x,y\rangle >1\},\ \ x\in
C.$$

Addition and scalar multiplication are jointly
continuous for the weak upper topology $w(D,C)$
so that $D$ with the weak upper topology is a topological
cone. Alternatively,
$D$ may be mapped into the product space $\oRp^C$ via the map $x\mapsto
\widehat x$. The weak upper topology $w(D,C)$ is the topology induced
by the upper topology on the product space $\oRp^C$.

As we suppose the bilinear form to be non-singular, the weak upper
topology $w(D,C)$ satisfies the T$_0$ separation axiom.
The specialization order is given by $y\leq y'$ if $\langle x,y\rangle \leq
\langle x,y'\rangle$ for all $x\in C$, that is, if $\widehat y(x)\leq
\widehat{y'}(x)$ for all $x\in C$. 

\begin{rem}
We may define a weak upper topology $w(D,B)$ for any subset $B$ by
taking the coarsest topology rendering continuous the maps
$\widehat x$ for all $x \in B$. This weak upper topology agrees with the
weak upper topology $w(D,\mathsf{Cone}(B))$, where $\cone(B)$ denotes
the subcone of $C$ generated by $B$. This follows from the fact that
finite linear combinations of lower semicontinuous functionals are
also lower semicontinuous.
\end{rem}

As the weak upper topology is generated by the subbasic open sets
$U_x$, $x\in C$, the finite intersections \[U_F = \{y\in D\mid \langle
x,y\rangle > 1 \mbox{ for all } x\in F\}\] where $F$ is a finite
subset of $C$ form a basis of the weak upper topology. 
Since these basic open sets are convex, this topology is locally
convex in the sense that every point has a neighborhood basis of open
convex sets.  
Every $w(D,C)$-open set is the union of a family of
basic open sets $U_F$. Translating this statement 
to the homogeneous functionals according to Lemma \ref{lem:hom} we
obtain\display  

\begin{lem} 
If $D$ carries the weak upper topology $w(D,C)$, the  homogeneous lower
semicontinuous functionals  $\psi\colon D\to\oRp$ are those  
obtained from the functionals $\widehat x$, $x\in C$, as pointwise
sups of finite pointwise infs. 
\end{lem}

We now come to our general form of Lemma \ref{lem:2}\display

\begin{thm}\label{main}
Let $C,D$ be a pair of cones together with a non-singular bilinear map $C\times
D\to\oRp$. We equip $D$ with the weak upper topology $w(D,C)$. Then
every lower semicontinuous sublinear functional $\psi\colon
D\to\oRp$ is the pointwise supremum of a family of point evaluations
$\widehat x$; more explicitly: For every $y\in D$, 
\[\psi(y) = 
\sup\{\widehat x(y)\mid x\in C , \widehat x\leq \psi\} 
= \sup\{\langle x,y\rangle\mid \forall z\in D.\ \langle x,z\rangle
\leq \psi(z)\}\]
\end{thm}

\proof
By the preceding lemma every homogeneous lower semicontinuous
functional $\psi$ on $D$ is the pointwise sup 
of finite pointwise meets $\widehat x_1\wedge\dots\wedge\widehat x_n$ where the
$x_i$ range over finite families of elements in $C$. If $\psi$ is
sublinear, Proposition \ref{key}
tells us that there is a convex combination $\sum_{i=1}^n
a_i\widehat x_i,\ a_i\in\Rp, \sum_i a_i =1$, such that $\widehat
x_1\wedge\dots\wedge\widehat x_n\leq  \sum_{i=1}^n a_i\widehat x_i\leq
\psi$. Since $\sum_{i=1}^n a_i\widehat x_i = \widehat{\sum_{i=1}^n
  a_ix_i}$, we have found an element 
$x= \sum_{i=1}^n a_ix_i$ in $C$ with $\widehat
x_1\wedge\dots\wedge\widehat x_n\leq \widehat x\leq\psi$. Thus
$\psi$ is the pointwise sup of functionals $\widehat x$.  
\qed 

Concerning the previous proof, note that the functional $\widehat
x_1\wedge\dots\wedge\widehat x_n$ is superlinear and lower
semicontinuous. Thus the Sandwich Theorem
\cite[8.2]{keimel08} allows to find a continuous linear functional
$\varphi$ with 
$\widehat x_1\wedge\dots\wedge\widehat x_n\leq \varphi\leq\psi$. But
that is not what we want. We
want to show that we can choose $\varphi=\widehat x$ for some $x\in  C$. For
this we use Proposition \ref{key} which was proved through a reduction of the problem 
to the finite dimensional case.

In the following we rephrase our theorem for some special situations.

\subsection{Dual pairs of ordered cones} 
One specializes the cones $C$ and
$D$ above by a dual pair of ordered cones $C$ and 
$D$ with an order preserving nonsingular bilinear map  $(x,y)\mapsto\langle
x,y\rangle\colon C\times D\to\oRp$. Note that the given order on $D$
may be stronger than the 
specialization preorder induced by the weak upper topology $w(D,C)$. Both
orders agree  if and only if the bilinear form is \emph{order 
non-singular} in the sense that for $x\not\leq x'$ in $C$ there is a
$y\in D$ such that $\langle x,y\rangle\not\leq \langle x',y\rangle$,
and similarly for the other argument. Under these hypotheses, Theorem \ref{main}
specializes to:

\begin{cor}
For every sublinear functional $\psi\colon D\to \oRp$ which is lower
semicontinuous with respect to the weak upper topology $w(D,C)$ we have
$\psi(y)=\sup\{\widehat x(y)\mid x\in C,\ \ \widehat x\leq \psi\}$ for
every $y\in D$.
\end{cor}

\subsection{Dual pairs of semitopological cones}

We now consider a pair $C,D$ of semitopological cones together with a
separately continuous non-singular bilinear form $(x,y)\mapsto\langle
x,y\rangle\colon C\times D\to\oRp$. The weak
upper topology $w(D,C)$ is coarser than the given topology on
$D$. Applying Theorem \ref{main} under these hypotheses we obtain:

\begin{cor}
For every sublinear functional $\psi\colon D\to \oRp$ which is
lower semicontinuous with respect to the weak upper topology $w(D,C)$ we
have $\psi(y)=\sup\{\langle x,y\rangle\mid x\in C,\ \ \widehat x\leq
\psi\}$ for every $y\in D$.  
\end{cor} 

Since the given topology on $D$ may be strictly coarser than the weak
upper topology $w(D,C)$, the claim of the
corollary need not be true 
for the sublinear functionals $\psi\colon D\to \oRp$ which are
lower semicontinuous with respect to the original topology on $D$.
 
\subsection{Dual pairs of d-cones}
Recall that a poset $D$ is \emph{directed complete} (a \emph{dcpo},
for short), if every directed family $(x_i)_i$ of elements in $D$ has a
least upper bound, denoted by $\dsup_i x_i$. A map $f\colon C\to D$
between dcpos is called Scott-continuous if it preserves least upper
bounds of directed sets, that is, if it is order preserving and if
$f(\dsup_i x_i)=\dsup_if(x_i)$ for every directed family of elements
in $C$. These are precisely the functions that are continuous for the
respective Scott topologies, where a subset $A$ of a dcpo is called
Scott-closed if $A$ is a lower set and if $\dsup_i x_i\in A$ for every
directed family of $x_i\in A$. We refer to \cite{gierz03} for
background on dcpos. 

A \emph{d-cone} is a cone equipped with directed complete partial order
in such a way that scalar multiplication $(r,x)\mapsto rx\colon\Rp\times
C\to C$ and addition $(x,y)\mapsto x+y\colon C\times C\to C$ are
Scott-continuous. With respect to the 
Scott topology a d-cone is a semitopological cone. $\oRp$ with its
usual order is a d-cone; its Scott topology agrees with the upper
topology. Thus the Scott-continuous functionals on a d-cone are the
lower semicontinuous ones.

We can specialize our results to dual pairs of 
d-cones $C,D$ with a Scott-continuous non-singular bilinear form $\langle  -,
-\rangle\colon C\times C\to C$. But here we want to consider also subcones
$C_0\subseteq C$ 
which are \emph{d-dense} in $C$, that is, the only d-subcone of $C$ that
contains $C_0$ is $C$ itself. We have\display

\begin{lem}\label{subbase}
The weak upper topology $w(D,C_0)$
agrees with the weak upper topology $w(D,C)$.
\end{lem}

\proof
In order to verify our
claim, we first notice: If $(x_i)_i$ is a directed family in $C$ and
$x=\dsup_ix_i$, then $U_{x}=\bigcup_i U_{x_i}$. Indeed, $y\in U_{x}$ if
and only if $1<\langle x,y\rangle =\langle \dsup_i x_i,y\rangle
=\dsup_i\langle x_i,y\rangle$ (by the Scott-continuity of the bilinear
form) if and only if 
$1<\langle x_i,y\rangle$ for some $i$ if and only if $y\in U_{x_i}$
for some $i$. This shows that adding suprema of directed sets to the
set $C_0$ does not refine the weak upper topology. By \cite[Corollary
6.7]{KL} this suffices to justify our claim.
\qed 

Under the hypotheses of this subsection, Theorem \ref{main} now
specializes to the following: 

\begin{cor}
Let $C,D$ be a dual pair of d-cones and $C_0$ a d-dense subcone of
$C$. We endow $D$ with the weak  
upper topology $w(D,C_0)$ and consider a Scott-continuous
sublinear functional $\psi\colon D\to \oRp$. Then
$\psi(y)=\sup\{\widehat x(y)\mid x\in C_0, \widehat x\leq \psi\}$ for
every $y\in D$.
\end{cor}

\section{Dual cones and the Schr\"oder-Simpson Theorem}

The standard situation for applying the previous setting 
is to start with a cone $C$, to form the \emph{algebraic dual} $D=C'$
of all linear 
functionals $\mu\colon C\to\oRp$ which is again a cone under pointwise
addition and scalar multiplication and to take the bilinear map
$\langle x,\mu\rangle =\mu(x)$ for $x\in C,\ \mu\in C'$ which is
non-singular for obvious reasons.  

For ordered cones $C$ it is natural to restrict to the \emph{order
  dual} of all order preserving linear functionals.

If we start with a
semitopological cone $C$, we consider 
the \emph{topological dual} $C^*$ of all lower semicontinuous linear
functionals on $C$ which is a subcone of the algebraic dual $C'$. We
 consider the weak upper topology $w(C^*,C)$ on $C^*$ which is 
also called the \emph{weak$^*$upper topology} in
agreement with the terminology in functional analysis. It is the
coarsest topology for which the evaluation maps  
\[\widehat x = (\mu\mapsto \mu(x))\colon C^*\to\oRp\]
are lower semicontinuous for all $x\in C$.
The bilinear map $(x,\mu)\mapsto \mu(x)\colon C\times 
C^*\to \oRp$ is separately lower semicontinuous. We ask\display

\begin{question}\label{qu:bij}
Characterize those semitopological cones $C$ with the property that,
for every linear functional $\varphi\colon C^*\to\oRp$ that is lower
semicontinuous for the weak$^*$upper topology,  there is a unique
element $x\in C$ such that $\varphi(\mu)=\mu(x)$ for every $\mu\in
C^*$, that is, $\varphi =\widehat x$ for a uniquely determined $x\in C$. 
\end{question}


Since the evaluation maps are linear, $x\mapsto \widehat x$
yields a map from $C$ to the double dual $C^{**}$ which is
linear. The question formulated above is equivalent
to the question whether the map $x\mapsto \widehat x\colon C\to
C^{**}$ is an isomorphism of cones.  

A simple example for cones having this property are the 'finite dimensional' 
cones $\oRp^n$ endowed with the upper
topology (which agrees with the product topology with respect to
the upper topology on $\oRp$). This follows 
from the fact that the  dual 
cone of all lower semicontinuous linear functionals on $\oRp^n$ is
isomorphic to $\oRp^n$ and its weak$^*$upper topology agrees with the
upper topology. For the proof we just need the following\display 

\begin{lem}\ref{ex:rn}
For every lower semicontinuous linear functional $G\colon \oRp^n\to
\oRp$ there are $r_1,\dots,r_n\in\oRp$ such that
$G(x_1,\dots,x_n)=r_1x_1+\dots+ r_nx_n$. 
\end{lem}

\proof
Given $G\colon\oRp^n\to\oRp$, let $r_i=G(e_i)\in \oRp$ where $e_i$ is the
standard basis vector all entries of which are zero, except the {i}-th
entry which is $1$. By linearity, $G(x_1,\dots,x_n)=\sum_i r_ix_i$ for
all $(x_1,\dots,x_n)\in\Rp^n$. By lower semicontinuity this formula
extends to all of $\oRp^n$.   
\qed

The Schr\"oder-Simpson Theorem yields a whole class of examples for
which the answer to Question \ref{qu:bij} is affirmative. Indeed, for
any topological space $X$, the cone
$C=\cL X$ of lower semicontinuous functions $f\colon X\to\oRp$ is a d-cone
and its dual is the valuation powerdomain $C^*=\cV X$.

 Of course,  Question \ref{qu:bij} can be split. 
%
The injectivity of the map $x\mapsto \widehat x$ is equivalent to the
question whether the 
lower semicontinuous linear functionals on $C$ separate the points of
$C$. Indeed, for elements $x\neq y$ in $C$, there is a $\mu\in C^*$
such that $\mu(x)\neq\mu(y)$ if, and only if, $\widehat x(\mu)\neq
\widehat y(\mu)$, that is, $\widehat x\neq \widehat y$. 
This criterion for injectivity is equivalent to the property that,
whenever $x\not\leq y$, there is a convex open set $U\subseteq C$
containing $x$ but not 
$y$. This is a consequence of the Separation Theorem
(\cite[9.1]{keimel08}). 
This property is in particular guaranteed by local convexity: A
semitopological cone is \emph{locally convex} if
every point has a neighborhood base of open convex sets. 
Thus, injectivity is not the problem;
surjectivity is the issue.

We are mainly interested in the case of a d-cone $C$ with its Scott
topology. We then form the dual cone $D=C^*$ of all 
Scott-continuous linear maps $\mu\colon C\to\oRp$ which is a
d-cone, too, for the pointwise defined order. The bilinear map
$\langle x,\mu\rangle =\mu(x)$ is Scott-continuous. 
Again, the topology $w(C^*,C)$ on $C^*$ is called the weak$^*$upper
topology. By Lemma \ref{subbase} above, we can replace $C$ by any
d-dense subcone $C_0$ since then $w(C^*,C)=w(C^*,C_0)$.
 Theorem \ref{main} now specializes to the following general version
of Lemma \ref{lem:2} in the Introduction\display

\begin{cor}
Let $C$ be a d-cone and $C^*$ its dual cone of Scott-continuous linear
functionals $\mu\colon C\to\oRp$ endowed with its weak$^*$-upper
topology. Let $C_0$ be a d-dense subcone of $C$ in the sense 
that there is no proper d-subcone of $C$ containing $C_0$. Then for
every lower semicontinuous sublinear functional $\varphi\colon C^*\to
\oRp$ we have $\varphi(\mu)= \sup\{\mu(x)\mid x\in C_0, \widehat x\leq
\varphi\}$ for every $\mu\in C^*$, that is, $\varphi=\sup \{\widehat
x\mid x\in C_0,\ \widehat x\leq \varphi\}$. 
\end{cor}

What we are lacking is a general version of Lemma \ref{lem:1}. So we
are left with the 
question: Characterize the d-cones $C$ for which the conclusion of the
previous corollary can be strengthened as follows\display
\begin{itemize}
\item[] For every lower semicontinuous linear functional $\varphi\colon
C^*\to\oRp$ there is a directed family $x_i$ of elements in $C$ such
that $\varphi(\mu)=\dsup\!\!\!_i\ \mu(x_i)$ for all $\mu\in C^*$. 
\end{itemize} 
If this is the case, then we may form $x=\dsup\!\!\!_i\ x_i$
and we obtain  $\varphi(\mu)=\dsup\!\!\!_i\
\mu(x_i)=\mu(\dsup\!\!\!_i\ x_i)=\mu(x)$,
since all the $\mu\in C^*$ are Scott-continuous.

\section{Comments}

The following example may be instructive. Even in the finite
dimensional case not every linear functional on a cone with values in
$\Rp$ may be lower semicontinuous for a weak upper topology.

\begin{exa}
 Take the discrete two elements space $2=\{0,1\}$ and the
  Sierpinski space $\Sigma$, the two element set $\{0,1\}$ with
  $\{1\}$ as the only proper nonempty open subset, let 
 $D=\cL 2 =\Rp^2$ and $C=\cL \Sigma=\{(x_1,x_2)\in\Rp^2\mid x_1\geq
 x_2\}$. The standard 
 inner product is a bilinear map $C\times D\to\Rp$. One might think
 that this situation gives a counterexample to Theorem
 \ref{main}. Indeed, the 
 second projection $\pi_2=((y_1,y_2)\mapsto y_2)\colon D\to\Rp$ is a
 linear functional, but 
there is no $x=(x_1,x_2)\in C$ such that $\pi_2(y_1,y_2) = \langle
 (x_1,x_2),(y_1,y_2)\rangle$. (The only point $x$ satisfying this equation
 is $x=(0,1)$ which does not belong to $C$.) Also, it is not possible
 to obtain $\pi_2$ as the pointwise sup of functionals of the type
 $\widehat x =(y\mapsto\langle x,y\rangle)$ with $x\in C$. But there
 is not contradiction 
 to Theorem \ref{main}. The point is, that $\pi_2$ is not lower
 semicontinuous with respect to the weak upper topology
 $w(D,C)$. Indeed, the specialization order $\leq_s$ associated with
 the topology $w(D,C)$ has to be preserved by functionals that are lower
 semicontinuous with respect to the weak upper topology $w(D,C)$. In
 this case the specialization order on $D$ is given by \[(y_1,y_2)\leq_s
 (y_1',y_2')\iff y_1\leq y_1',\ y_1+y_2\leq y_1'+y_2'  \]    
The projection $\pi_2$ does not preserve $\leq_s$, since $(1,1)\leq
(2,0)$ but $\pi_2(1,1)=1 \not\leq 0 =\pi_2(2,0)$.
\end{exa}

In the following we show a phenomenon that looks surprising\display

\begin{exa}
\emph{Let $C$ and $D$ be subcones of $\R^n$ which are closed and
have interior points for the usual Euclidean topology. Suppose in
addition that, for the standard inner product, $\langle c,d\rangle
\geq 0$ for all $c\in C, d\in D$. We claim: $C$ is the dual $D^*$ of
$D$ in the sense that for every 
 linear functional $\varphi\colon D\to\Rp$, which is lower
 semicontinuous for the 
 weak upper topology $w(D,C)$, there is a unique $x\in C$ such that
 $\varphi=\widehat x$ for some  $x\in C$.}

For the proof consider such a linear functional $\varphi$. It may be
 extended in a unique way to a
 linear functional on $\R^n$. Thus there is a unique element
 $x_0\in\R^n$ such that $\varphi=\widehat{x_0}$. By Proposition
 \ref{main}, $\langle 
 x_0,y\rangle =\sup_{v\in X}\langle v,y\rangle$ for all $y\in D$, where
 $X$ is the set of all $v\in C$ such that $\widehat v\leq \widehat{x_0}$,
 that is $\langle v,y\rangle\leq \langle x_0,y\rangle$, that is $\langle
 x_0-v,y\rangle\geq 0$ for all $y\in D$. This amounts to say that
 $x_0-v\in D^*$, that is $v\leq_{D^*} x$ for the order $\leq_{D^*}$
 defined by the cone $D^*$. This shows that the set $X$ is
 compact. Clearly, $X$ is  
 convex. If $x_0\not\in C$, then $x_0-X=\{x_0-x\mid x\in X\}$ is
 compact and convex but 
 does not contain $0$.   
Take any $y_0$ in the interior of $D$. Then $\widehat{y_0}$ is a
strictly positive functional on $D^*$, that is, $\langle x,y_0\rangle>
0$ for all $x\neq 0$ in $D^*$. Hence, $\langle x-v,y_0\rangle>0$ for
all 
 $v\in X$. Since $X$ is compact, it follows that $\sup_{v\in X}\langle
 v,y_0\rangle <  \langle x,y_0\rangle$, a contradiction. \\

Similar results as above hold for d-subcones of $\oRp^n$. But note
that $C$ is not the dual of $D$, in general, if we use the usual
Euclidean topology on $D$.
\end{exa}


\begin{thebibliography}{}

\bibitem{AB}
\newblock C. D. Aliprantis and K. C. Border,
\newblock Infinite Dimensional Analysis: A Hitchhiker's Guide.
\newblock 3$^{rd}$d edition, Springer Verlag (2006).

\bibitem{BSS}
\newblock I. Battenfeld, M. Schr\"oder and A. Simpson,
\newblock A convenient category of domains.
In: Computation, Meaning and Logic, Articles dedicated to Gordon Plotkin
L. Cardelli, M. Fiore and G. Winskel (eds),
Electronic Notes in Computer Science 172  (2007), pages 69--99. 

\bibitem{B}
\newblock K. C. Border,
\newblock Separating Hyperplane Theorems.
\newblock Notes, California Institute of Technology.\\ 
\url{www.hss.caltech.edu/~kcb/Notes/SeparatingHyperplane.pdf}. 


\bibitem{gierz03}
\newblock G.~Gierz, K.~H.~Hofmann, K.~Keimel, J.~D.~Lawson,
M.~Mislove, and D.~S.~Scott,
\newblock Continuous Lattices and Domains.
\newblock \emph{Encyclopedia of Mathematics and its Applications,
  vol. 93},
\newblock Cambridge University Press, 2003, xxxvi+591 pages.

\bibitem{GL}
\newblock J. Goubault-Larrecq,
\newblock A short proof of the Schr\"oder-Simpson theorem.
\newblock \emph{Mathematical Structures in Computer Science} 25 (2015), pages
1--5.

\bibitem{GL12}
\newblock J. Goubault-Larrecq,
\newblock Isomorphism Theorems between Models of Mixed Choice, 
\newblock Draft.

\bibitem{J}
\newblock C. Jones,
\newblock \emph{Probabilistic non-determinism}. PhD Thesis,
  Department of Computer Science, The University of Edinburgh (1990),
  201 pages (also published as Technical Report No.~CST-63-90).


\bibitem{keimel08}
K. Keimel,
\newblock Topological Cones: Functional analysis in a T$_0$-setting.
\newblock \emph{Semigroup Forum} 77 (2008), pages 108--142. 

\bibitem{keimel12}
K. Keimel,
\newblock Locally convex cones and the Schr\"oder-Simpson Theorem.
\newblock \emph{Quaestiones Mathematicae} 35 (2012), pages 353--390. 

\bibitem{KL}
K. Keimel and J. D. Lawson,
\newblock Extending algebraic operations to D-completions.
\newblock \emph{Theoretical Computer Science} 430 (2012), pages 73--87.

\bibitem{Ki}
O. Kirch,
\newblock Bereiche und Bewertungen.
\newblock MSc Thesis, TU Darmstadt, 1993.\\
\newblock \url{http://www.mathematik.tu-darmstadt.de/fbereiche/logik/research/Domains/Domains.html}


\bibitem{Plo} G. D. Plotkin, 
\newblock A domain-theoretic Banach-Alaoglu theorem. 
\newblock \emph{Mathematical Structures in Computer Science}
\textbf{16} (2006), pages 299--312.

\bibitem{Ru} W. Rudin, 
\newblock Functional Analysis. 
\newblock McGraw-Hill Book Company, 1973.


\bibitem{Sch}
H.~H.~Schaefer,
\newblock Topological Vector Spaces.
\newblock Springer Verlag, 3rd printing (1970).  

\bibitem{SS}
M. Schr\"oder and A. Simpson,
\newblock Probabilistic observations and valuations. 
\newblock \emph{Electronic Notes in Theoretical Computer Science}  \textbf{155} (2006), 605--615.


\bibitem{SS1}
M. Schr\"oder and A. Simpson,
\newblock Probabilistic Observations and Valuations.
\newblock Talk at New Interactions Between Analysis, Topology and Computation,
Birmingham (2009).
\newblock \url{http://homepages.inf.ed.ac.uk/als/Talks/birmingham09.pdf}

\end{thebibliography}
\end{document}